\documentclass[a4paper]{jpconf}
\usepackage{graphicx}
\begin{document}
\title{Galaxy rotation curves from a fourth order gravity}

\author{Priti Mishra and Tejinder P Singh}

\address{Tata Institute of Fundamental Research, Homi Bhabha Road, Mumbai, India}

\ead{priti@tifr.res.in, tpsingh@tifr.res.in}

\begin{abstract}
While the standard and most popular explanation for the flatness of galaxy rotation curves 
is dark matter, one cannot at this stage rule out an explanation based on a modified law of 
gravitation, which agrees with Newtonian gravitation on the scale of the solar system, but 
differs from it on larger length scales. Examples include Modfied Newtonian Dynamics [MOND] 
and Scalar-Tensor-Vector Gravity [STVG]. Here we report on a fourth order modification of the 
Poisson equation which yields the same Yukawa type modification of Newtonian gravity as STVG, 
and which can explain flat galaxy rotation curves for a large sample of galaxies, once specific 
values for two parameters have been chosen. We speculate on two possible origins for this 
modified Poisson equation: first, a possible fourth order modification of general relativity, 
and second, quadrupole gravitational polarization induced on a galaxy because of the pull of neighbouring galaxies.   
\end{abstract}

\section{Introduction}
Flat galaxy rotation curves and the departure from the expected Keplerian fall-off which they 
imply are most popularly attributed to the presence of cold dark matter [CDM] in the galaxy. 
Together with a cosmological constant $\Lambda$, CDM constitutes the highly successful 
$\Lambda$CDM model which explains most of cosmological, cluster and galaxy data. 
Nonetheless, considering that dark matter has not yet been detected in the laboratory, 
one should tentatively also allow for the possibility that flat galaxy rotation curves 
arise, not from the presence of dark matter, but from a possible modification of the 
law of gravitation on scales larger than the solar system. 

Notable examples of such modifications include Modified Newtonian Dynamics [MOND] ~\cite{MOND}
where it is assumed that the gravitational force acting on a test particle of mass $m$ is  given by the relation
\begin{equation}
F=m a\mu\bigg(\frac{a}{a_0}\bigg)
\end{equation}
where $a$ is acceleration in Newtonian mechanics and $a_0$ is a new fundamental constant of nature having the value 
$2\times 10^{-10}$ ms$^{-2}$. 
For very small accelerations [large distances] it is assumed that 
\begin{equation}
\mu\bigg(\frac{a}{a_0}\bigg)=\frac{a}{a_0}.
\end{equation}
whereas $\mu$ approaches the value one for accelerations encountered in the solar neighborhood. 

Hence for large distances we will have the relation 
\begin{equation}
\frac{GM}{r^2}=\frac{a^2}{a_0}\\
\Rightarrow a=\frac{\sqrt{GMa_0}}{r}.\label{MONDacceleration}
\end{equation} 
The virtue thus is that $a$ falls as $1/r$ rather than $1/r^2$. Thus equating $a$ to the 
centripetal acceleration $v^2/r$ we get that
\begin{equation}
\frac{v^2}{r}=a=\frac{\sqrt{GMa_0}}{r}\quad \quad \\
\Rightarrow\quad  v=(GMa_0)^{1/4}.
\end{equation}
With the numerical choice of $a_0$ made above one gets the desired flat value of the velocity.

Another noted example is Scalar-Tensor-Vector Gravity [STVG], where a massive vector 
field with a mass, denoted $\mu$, interacts with gravity with a coupling strength denoted 
$\omega$ ~\cite{Moffat}. It can be shown that in STVG, the law of gravitation acquires a Yukawa modification, which is given by
\begin{equation}
\label{accelerationlaw} a(r)=-\frac{G
M}{r^2}\biggl\{1+\sqrt{\frac{M_0}{M}}\biggl[1-\exp(-r/r_0)
\biggl(1+\frac{r}{r_0}\biggr)\biggr]\biggr\}.
\end{equation}
Here, $r_0=1/\mu$ and $M_0$ is a parameter which vanishes when $\omega=0$.
It is assumed [and justified by further considerations] that one can generalize this to the case of a mass distribution by
replacing the factor $GM/r^2$ in (\ref{accelerationlaw}) by
$GM(r)/r^2$. The rotational velocity of a star $v_c$ is
obtained from $v_c^2(r)/r=a(r)$ and is given by
\begin{equation}
v_c=\sqrt{\frac{G
M(r)}{r}}\biggl\{1+\sqrt{\frac{M_0}{M}}\biggl[1-\exp(-r/r_0)
\biggl(1+\frac{r}{r_0}\biggr)\biggr]\biggr\}^{1/2}.
\label{rotvel}
\end{equation}

From observations of galaxies the density profile $\mu(r)$ is known to be 
\begin{equation}
\mu(r)=\frac{3}{4\pi r^{3}} \beta  M(r) \bigg[\frac{r_c}{r+r_c}\bigg]
\label{densprof}
\end{equation}
where
\begin{equation}
M(r)=4\pi \int_0^rdr'r'^2\mu(r') = M \left( \frac{r}{r+r_{c}}\right)^{3\beta}
\end{equation}
and
\begin{equation}
\beta = \left\{
\begin{array}{ll} 1 & \mbox{for HSB galaxies,} \\
2 & \mbox{for LSB \& Dwarf galaxies.}
\end{array} \right.
\end{equation}

 A good fit to a large number of galaxies has been
achieved with the parameters:
\begin{equation}
M_0=9.60\times 10^{11}\,M_{\odot},\quad r_0=13.92\,{\rm
kpc}=4.30\times 10^{22}\,{\rm cm}.
\end{equation}
In the fitting of the galaxy rotation curves for both LSB and HSB
galaxies, using photometric data to determine the mass
distribution $M(r)$,  only the
mass-to-light ratio $\langle M/L\rangle$ is employed, once the
values of $M_0$ and $r_0$ are fixed universally for all LSB and
HSB galaxies. Dwarf galaxies are also fitted with the
parameters
\begin{equation}
M_0=2.40\times 10^{11}\,M_{\odot},\quad r_0=6.96\,{\rm
kpc}=2.15\times 10^{22}\,{\rm cm}.
\end{equation}

\section{Fourth order gravity}
Remarkably enough, it can be shown that the modified law of acceleration (\ref{accelerationlaw}) 
of STVG,
and the consequent velocity profile (\ref{rotvel}) is also a solution of the following modified 
fourth order Poisson equation ~\cite{priti-tp}
\begin{equation}
\nabla^4\phi-k^2\nabla^2\phi=-4\pi Gk^2\mu(r)
\label{fog}
\end{equation}
provided the density profile is chosen as in (\ref{densprof}) and one identifies $k=1/r_0$. 
$M_0$ appears as a constant of integration in the solution of (\ref{fog}). 

Greater insight into the modified acceleration can be had by expanding the acceleration law
\begin{equation}
a(r)=-\frac{GM(r)}{r^2}\bigg\{1+\sqrt{\frac{M_0}{M}}\bigg[1-\exp(-r/L)\bigg(1+\frac{r}{L}\bigg)\bigg] \bigg\}
\label{yukawa}
\end{equation}
 around $r=r_0$ and writing it as a sum of two terms : a part that falls as $1/r^2$ and 
is independent of $r_0$, and a part that falls as $1/r$ and depends on $r_0$. This gives

\begin{equation}
a(r)\approx - \frac{GM(r)}{r^2}\bigg[1+\sqrt{\frac{M_0}{M}}\bigg\{1-\frac{3}{e}\bigg\}\bigg]- \frac{GM(r)}{r}\bigg[
\sqrt{\frac{M_0}{M}}\frac{1}{r_0 e}\bigg].
\label{accelrationintwoparts}
\end{equation}

\begin{figure}[!ht]
\begin{center}
\includegraphics[width=\textwidth]{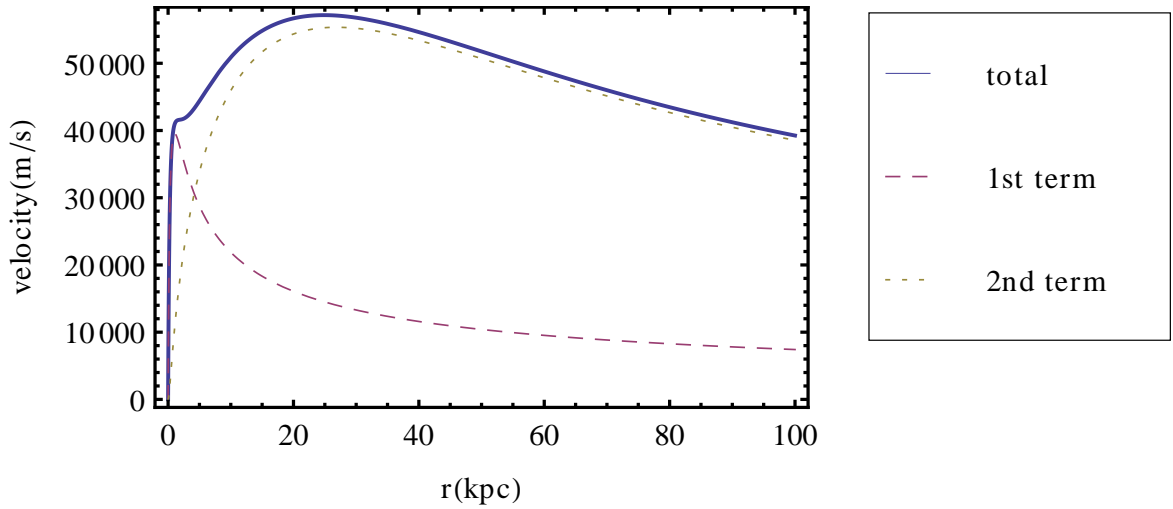}
\caption{Dashed curve: velocity due to first term, dotted curve: velocity due to second term, 
thick curve: total velocity.}
\label{total velocity}
\end{center}
\end{figure}

The second term dominates for $r\geq r_0$. Fig. 1 above plots the rotation velocity curve, 
and it is clear that the Keplerian fall-off due to conventional Newtonian acceleration 
[first term] is modified to a flat rotation curve [second term] for $r\geq r_0$. The 
$1/r$ fall-off brought about by the second term is MOND-like behaviour. At even larger 
distances, the second term is exponentially damped and the fall off is again Keplerian 
but with an effectively larger value of $G$.

It is significant that from our work we can give an estimate of the theoretical value of 
$a_0$ in MOND. A simplistic guess would be to construct a quantity 
with the dimension of acceleration from our fundamental quantities $k$ and $M_0$. This is 
nothing but 
\begin{eqnarray}
GM_0k^2=\frac{GM_0}{r_0^2}=&=&\frac{6.67\times 10^{-11}\times 9.60\times10^{11}\times 
2\times 10^{30}}{(13.92\times 10^3\times3.08\times 10^{16})^2}\\
&=&3\times 10^{-10} \ \rm{ms^{-2}}
\end{eqnarray}
which is of the same order as $a_0$ in MOND. ( $a_0=2\times 10^{-10}$m/s$^2$). The relation
 is made more transparent by comparing the 
acceleration 
given by the second term of Eqn. (\ref{accelrationintwoparts}) with the acceleration in 
MOND in Eqn. (\ref{MONDacceleration}). This comparison 
yields the fundamental relation 
\begin{equation}
a_0=\frac{GM_0}{r_0^2e^2}
\end{equation}

\section{Possible origin of fourth order gravity} 

The fourth order modified Poisson equation considered above is the weak field limit of the 
following modification of general relativity:
\begin{equation}
R^{\mu\nu} - \frac{1}{2} g^{\mu\nu} R = \frac{8\pi G}{c^4} T^{\mu\nu} + 
k^{-2} R^{\mu\nu\alpha\beta}_{\ \ \ \ \ ;\alpha\beta}
\label{modee}
\end{equation}
It is possible that such an effective modification of Einstein gravity comes into play during the late 
stages of the evolution of the Universe, when large scale structures form.

An intriguing alternate possibility, which has been discussed in ~\cite{priti-tp}  is that the fourth 
order modification of the Poisson equation is a consequence of induced polarization of the gravitational 
field of a galaxy, caused by neighboring galaxies. Thinking of galaxies as `molecules' made of atoms 
[the stars], one would like to analyze how the averaged gravitational field inside a galaxy, is modified 
by the polarization of the molecules, due to the external pull of other galaxies.  Indeed one has at the 
back of the mind the polarization induced modification of electromagnetic fields in an electrically charged medium.

It can be shown that as a consequence of polarization, the averaged Einstein equations are effectively modified to
\begin{equation}
R_{\mu\nu}^{(0)}-\frac{1}{2}g_{\mu\nu}^{(0)}R^{(0)}=-\kappa(T_{\mu\nu}^{(free)}+\frac{1}{2}c^2
Q_{\mu\rho\nu\sigma}^{;\rho\sigma}),
\label{einstein-molecule}
\end{equation}
where  $T_{\mu\nu}^{(free)}$ is the energy-momentum 
tensor of molecules, and  $Q_{\mu\rho\nu\sigma}$ is the quadrupole gravitational polarization tensor.
It is modelled by assuming it be proportional to the Weyl tensor, in analogy with the electrostatic case, 
\begin{equation}
Q_{i0j0} = \epsilon_{g}C_{i0j0} 
\end{equation} and 
the gravitational dielectric constant $\epsilon _{g}$ is modelled by
\cite{Szekeres} as 
\begin{equation} 
\epsilon_g=\frac{1}{4}\frac{mA^2c^2}{\omega_0^2}N.
\end{equation}
Here, $A$ is the typical linear dimension of a molecule, $m$ is the typical mass of a molecule,
 $\omega_{0}$ is a typical frequency of harmonically oscillating atoms in the molecule and $N$ 
is the number density of the molecules in the medium. In the weak field limit, Eqn. (\ref{einstein-molecule}) 
can be shown to go over to the modified Poisson equation (\ref{fog}) with $k^2=c^2 / 2\pi G \epsilon_g$
~\cite{zala}. Furthermore, the theoretically predicted values of $k$ and $M_0$ agree well with the values
 phenomenologically assumed for fitting data, suggesting that gravitational quadrupole polarization is a 
contender for explaining flat galaxy rotation curves ~\cite{priti-tp}.


\section*{References}
\begin{thebibliography}{9}
\bibitem{MOND} Bekenstein J D 2006 {\it Contemporary Physics} {\bf 47} 387
\bibitem{Moffat} Brownstein J D and Moffat J W 2006 {\it Ap J} {\bf 636} 721
\bibitem{priti-tp} Mishra Priti and Singh T P 2011 arXiv:1108.2375 [astro-ph.GA] 
\bibitem{Szekeres} Szekeres P 1971 {\it Ann. Phys.} {\bf 64} 599
\bibitem{zala} Montani G, Ruffini R and Zalaletdinov R 2003 {\it Class. Quantum Grav.} {\bf 20} 4195.
\end{thebibliography}
\end{document}